# Surrogate Parenthood: Protected and Informative Graphs

Barbara Blaustein, Adriane Chapman, Len Seligman, M. David Allen, Arnon Rosenthal

The MITRE Corporation
7515 Colshire Drive
Mclean, VA 22102

{bblaustein, achapman, seligman, dmallen, arnie}@mitre.org

## ABSTRACT

Many applications, including provenance and some analyses of social networks, require path-based queries over graph-structured data. When these graphs contain sensitive information, paths may be broken, resulting in uninformative query results. This paper presents innovative techniques that give users more informative graph query results; the techniques leverage a common industry practice of providing what we call surrogates: alternate, less sensitive versions of nodes and edges releasable to a broader community. We describe techniques for interposing surrogate nodes and edges to protect sensitive graph components, while maximizing graph connectivity and giving users as much information as possible. In this work, we formalize the problem of creating a protected account G' of a graph G. We provide a utility measure to compare the informativeness of alternate protected accounts and an opacity measure for protected accounts, which indicates the likelihood that an attacker can recreate the topology of the original graph from the protected account. We provide an algorithm to create a maximally useful protected account of a sensitive graph, and show through evaluation with the PLUS prototype that using surrogates and protected accounts adds value for the user, with no significant impact on the time required to generate results for graph queries.

## 1. INTRODUCTION

Many graph-structured data sets contain sensitive information (e.g., a "secret source" in a provenance graph, a "member of gang x" relationship in a social network, or certain connections among computers or other entities in a network). Prior work has emphasized manipulations of the whole graph in order to permit analysis while still protecting node identity [1] or sensitive relationships [15]. Techniques include adding and removing edges based on knowledge of certain nodes' degree [10] or their neighborhood [16], sampling from a model of the network structure [9], and grouping entities into classes and masking the mapping between entities and corresponding nodes [5]. Such whole graph manipulations are appropriate when mining the graph for interesting patterns and trends.

However, there exist applications in which 1) only particular nodes, node properties, or relationship instances are sensitive, and 2) important queries traverse paths from specified starting places. For instance, in a provenance[1] graph, the *input-to* relationship is sensitive only in specific cases and not generically; the most common queries traverse paths from given starting points ("what data and processes contributed to this data?").

Figure 1 contains an abstract graph with some sensitive data instances (1a) and a partially ordered set of privilege classes (1b). (The shading in 1a indicates the minimum privilege required to access the node.) For our running discussions, we will interpret this graph as a social network. However, it could also be a provenance graph in which some data or edges are sensitive. (See Appendix Figure 11 for an extended provenance example similar to the abstract graph and roles in Figure 1.) It could also be a computer network in which the company owning the network may make connections among some nodes visible to anyone, while others are shared only with a newly acquired company ('High-1" privileges) or business partners ("High-2").

Using Figure 1 as a social network, in which nodes *c* and *g* represent individuals connected by the sensitive node *f*, representing involvement with a particular gang, a user may be willing to share that *c* and *g* know each other as long as the participation in the gang is hidden. However, there is currently no way to let a user with "High-2" privileges know that *c* and *g* are related.

Standard access controls require that only the information in Figure 1c be accessible to a user with "High-2" privileges. Unfortunately, for a path traversal query, if one ancestor node is too sensitive to access, one cannot return any of its predecessors—e.g., a "High-2" user seeking the predecessors of node *g* would learn nothing of *b* or *c*, despite the fact that they contain no sensitive information. It is not uncommon for the hiding of just a few nodes to cause most of a path to become inaccessible.

Our proposed solution leverages existing common practice; real-world data providers often create alternate, less sensitive versions of information that are releasable to different 1 "provenance of objects…is represented by an annotated causality graph, which is a directed acyclic graph, enriched with annotations capturing further information pertaining to execution." [11]

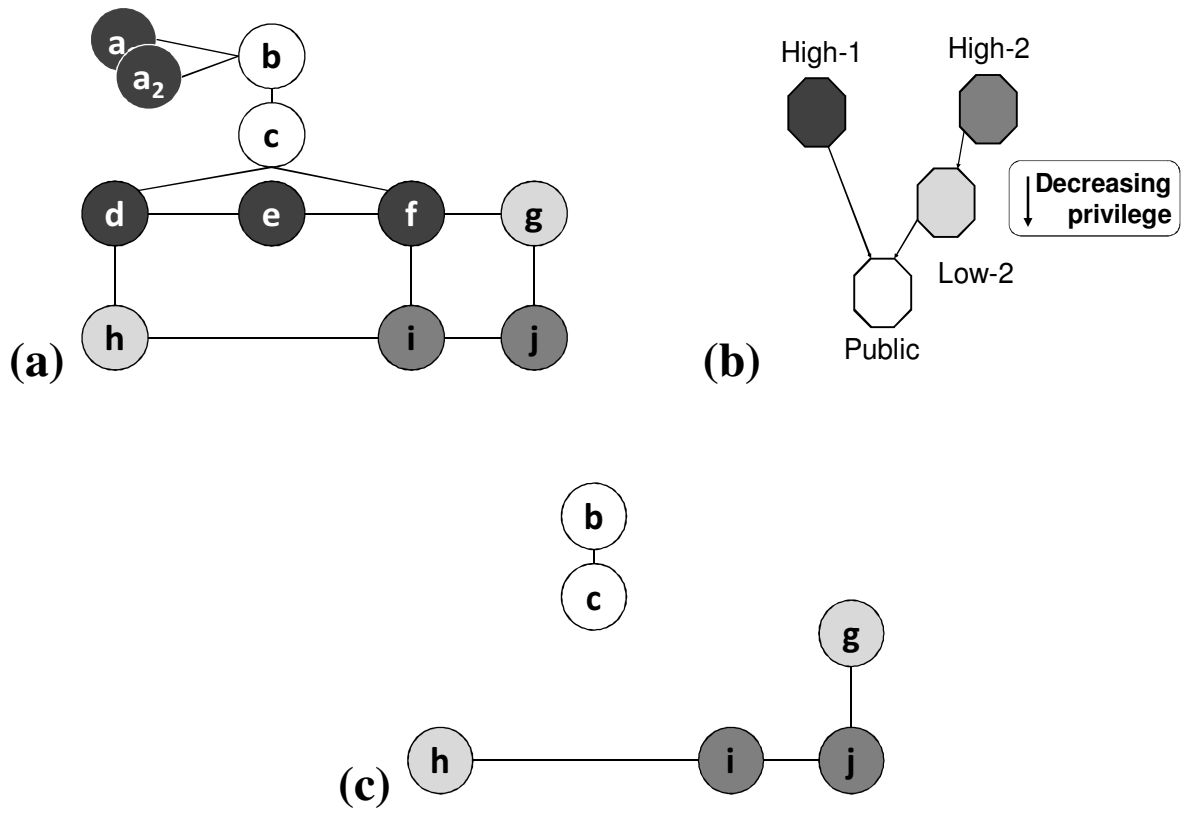

**Figure 1: (a) Sample Graph, $G$. This generic graph is used to illustrate provenance, social networking and computer networking examples in the introduction. It is also the basis for a running social network example throughout the rest of the paper. Nodes in graphs are shaded to reflect the lowest set of access privileges required. (b) Sample set of user privileges; each nickname ("High-1", etc.) summarizes a predicate expressing privileges. (c) $G_1'$, a naïvely protected account of 1(a) available to a user with "High-2" privileges.**

consumers. In the rest of this paper, we interpret Figure 1a as a social network used in a criminal investigation. It is possible that certain information ("court-sanctioned surveillance by an agent of Agency X") should only be revealed to highly trusted partners, say those with "High-2" privileges, while the larger set of "Low-2" partners would still get value from less detailed information ("a trusted law enforcement source"). We call such less sensitive versions *surrogates*. This paper describes techniques for interposing surrogate nodes and edges as needed to protect sensitive graph components, while maximizing graph connectivity and giving users as much information as possible. This paper makes the following contributions:

- We formalize the problem of creating a protected account $G'$ of a graph $G$, which does not include the sensitive information from $G$ while preserving as much connectivity as possible.
- We define a *utility* measure for protected accounts which allows comparisons of the informativeness of alternate protected accounts.
- We define an *opacity* measure for protected accounts, which indicates the likelihood that an attacker can recreate the topology of the original graph from the protected account.
- We present an algorithm for creating a maximally useful protected account from a graph with surrogates.
- We show, through evaluation with the PLUS prototype software [4], that using surrogates and protected accounts adds value for the user, with no significant impact on the time required to generate results for graph queries.

# 2. PRELIMINARIES

A graph $G$ consists of a set of nodes, $N$, and a set $E$ of directed edges between pairs of those nodes. (We model bi-directional edges as two directed edges.) Nodes have features, such as timestamp, author, etc., modeled as attribute-value pairs. A graph *object* is a node or edge appearing in that graph.

The provider of a graph object decides which information details may be released to consumers with specific privileges, e.g. "High-2". A provider can control the release of specific node features and, independently, control which incident edges a consumer may see [12]. For example, the social network may show two individuals as being business partners, but not reveal that one of them is the brother of a gang leader. We assume the existence of a Boolean function $authorized(c, o)$, for a consumer c and graph object o; this is evaluated by the cognizant authority for the object, using all context information (time, location, role, etc). It is out of the scope of this paper to explore ways in which consumer credentials are generated, passed, and authenticated.

To characterize which objects may be released to a particular consumer, we define privilege-predicates, Boolean functions over consumer credentials. A privilege-predicate $p$ defines a class of consumers:

*Definition 1. For a consumer c, a privilege-predicate p, and a graph object o, if p(c) ⇒ authorized(c, o), we say o is visible via p.*

For ease of reading this paper, we use a nickname for a particular privilege-predicate; in Fig. 1, for example, "High-1" refers to a specific privilege-predicate.

Privilege-predicates are partially ordered.

*Definition 2. A privilege-predicate $p_1$ <u>dominates</u> another privilege-predicate $p_2$ if ∀ consumers c, $p_1(c) \Rightarrow p_2(c)$. Note that a privilege-predicate trivially dominates itself.*

*Definition 3. A <u>lowest</u> predicate for an object o, $lowest(o)$, is a privilege-predicate $p$ such that $o$ is visible via $p$, and for all privilege-predicates $p'$, if $o$ is visible via $p'$, then $p'$ dominates $p$.*

A set of incomparable privilege-predicates is one in which none dominates another. In this paper, we assume that there is a "Public" privilege-predicate dominated by all other privilege-predicates.

# 3. NODE AND EDGE SURROGATES

Our goal is to provide each consumer with the most informative protected account possible, subject to the principle that it is more important to protect than inform.

## 3.1 Surrogate Nodes

Nodes in a graph may not be releasable to certain consumers, but there may be versions of these nodes releasable to other, less privileged, groups of consumers. These alternate versions, or *surrogate nodes*, protect sensitive information by omitting or changing features of the original node (e.g., using "illegal substance" instead of "heroin") for consumers lacking access to the original node: *If $n'$ is a surrogate for a node $n$, then $lowest(n')$ does not dominate $lowest(n)$.* (Note that this requirement allows $lowest(n')$ to be incomparable with $lowest(n)$.)

The node provider is responsible for generating appropriate surrogates and identifying the lowest privilege for each surrogate, although there is no requirement that surrogates be provided for every node. We define a <null> surrogate node with no features; <null> can be used as a default surrogate.

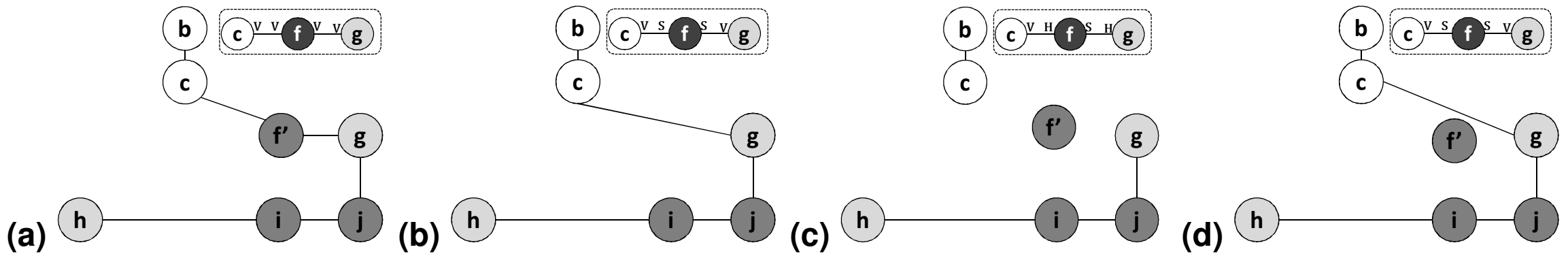

**Figure 2: Protected accounts with high-water set {"High-2"} based on Figure 1; relevant edge markings for "High-2" are shown in the dotted boxes for each graph. Assume all other edge markings are Visible. (a) Protected account with surrogate node *f'* and visible edges. (b) Protected account with hidden node *f* and surrogate edge *c-g*. (c) Protected account with surrogate node *f'* and hidden edges. (d) Protected account with surrogate node *f'* and same surrogate edge as graph in (b).**

*Definition 4. A node $n'$ corresponds to a node $n$ iff either all of the features of $n'$ are the same as those of $n$ (written $n' = n$) or $n'$ is defined as a surrogate for $n$).*

*Definition 5. Given a graph $G = (N, E)$, a protected account of $G$ is a graph $G' = (N', E')$ where each node $n' \in N'$ corresponds to a unique node $n \in N$ and for every path between two nodes in $G'$ there is a path in $G$ between the corresponding nodes.*

Thus, in Figure 1, each node in $G_1'$, the naïvely protected account, is the same as a single node in $G$, and all paths in $G_1'$ are also present in $G$.

For this work, perhaps the most fundamental characterization of a graph is the set of privilege-predicates needed to see all nodes. We define a *high-water set*, *HW*, that is the set of lowest privileges required to see all objects within a graph. In Figure 2a, the high-water set is {"High-1", "High-2"}. To see this complete graph, a consumer's credentials would have to satisfy a predicate dominating the conjunction of the members of the high-water set.

*Definition 6. The high-water set, $HW$, of a graph $G = (N, E)$ is the set of privilege-predicates $\{p_1, p_2 \ldots p_n\}$ such that*

*1. No member of* HW *dominates another,*

*2. For every node $n \in N$, there is some* p ∈ HW *such that* p *dominates* lowest(n)*, and*

*3. For every* p ∈ HW*, there is some node $n \in N$ such that* p = lowest(n).

While high-water sets are descriptive, they also provide a target when generating protected accounts. We would like to find a protected account whose high-water set comes as close as possible to the consumer's privileges while remaining dominated by them. If the consumer's privileges are identical to the lowest necessary for a given account, then that account contains nodes at the dominant levels allowable. Original nodes that are dominated by the consumer's privilege will appear in the protected account; original nodes which are not dominated by the protected account's high-water predicate will be omitted or represented by surrogate nodes (including the default <null>, if necessary) in the protected account.

## 3.2 Edges in Protected Accounts

Producers must be allowed to do more than just provide node surrogates. Relationships among nodes may also be sensitive, so edges in a protected account are also governed by providers' release policies. For instance, in the social network example, it is desirable to show the relationship between *c* and *g*, but omit the information that the path goes through a political cause.

We allow providers to mark all edges connected to a node (in practice, these may be defined on sets of nodes, e.g., all edges from data nodes of certain types, or all outgoing edges). Providers of information represented by each of the source and destination nodes may mark an edge. Markings at each end of an edge need not agree; this preserves local autonomy in the definition of release policies. Because each edge is subject to marking by at least the providers of the source and destination nodes, the final disposition of the edge in the protected account depends on the combination of markings.

*Definition 7. For a graph $G$ and a privilege-predicate $p$, each node-edge incidence has a marking, mark(n,e,p)∈{Visible, Hide, Surrogate}.*

A *Visible* marking means that the provider is willing to show this node-edge incidence to consumers satisfying *p*. *Hide* means it may not be shown, nor may it be used to compute any edge in the protected account. A *Surrogate* marking means that the node-edge incidence may be used to maintain a path in a protected account, although the incidence cannot be shown directly. The *Surrogate* marking is a way to hide the role of a node while maintaining some of the connectivity of the original graph.

*Surrogate edges* in a protected account $G'$ summarize paths in the original graph (e.g., an edge between *c* and *g*) that are visible via $HW_{G'}$.

*Definition 8. Given a high-water set $HW_{G'}$, and an original graph $G$, an HW-permitted path between nodes $n_1$ and $n_2$ in $G$ is one in which:*
*1. There are no edge-incidences marked Hide for any $p$ dominating a member of $HW_{G'}$ and in which the $n_1$ and $n_2$ edge-incidences are marked Visible for some $p$ dominated by a member of $HW_{G'}$, and*
*2. If there is a corresponding edge between $n_1$ and $n_2$ in $G$, then each node-edge incidence is marked Visible for some $p$ dominated by a member of HW.*

So, given the edge markings in Figure 2a, *c-f-g* is an HW-permitted path. Figure 2b shows a protected account with the

$$Path\ Utility\ Measure = \frac{\sum_{i=1}^{k} \%P(n')}{|N|} \qquad \frac{\left(\frac{1}{10}\right)+\left(\frac{1}{10}\right)+4\left(\frac{3}{10}\right)}{11} = .13 \qquad Node\ Utility\ Measure = \frac{\sum_{i=1}^{k} infoScore(n')}{|N|}$$

(a) (b) (c)

**Figure 3: Utility Measures**

surrogate edge *c-g*; showing a relationship without divulging information about the political cause.

# 4. UTILITY AND OPACITY MEASURES

This paper focuses on transforming a graph $G$ into a protected account $G'$, such that $G'$ preserves as much of the information in $G$, its *utility,* as possible, while protecting against attacks by malicious (or overly curious) users who try to use $G'$ to infer protected information in $G$, i.e., preserving the protected account's *opacity.*

## 4.1 Utility

Our goal is to generate protected accounts that convey more information than naïvely protected accounts. Connectedness is one measure of the information content of the graph, as captured in the Path Utility Measure below. Since we wish to support graph queries over paths, e.g. connections to or from specific nodes, we measure the percentage of paths retained in $G'$ to and from each node. For each node $n' \in G'$, the *path percentage*, $\%P(n')$, is the number of nodes in $G'$ connected (by any length path) to $n'$, divided by the number of nodes in $G$ that are connected to $n$. Recall from our definition of a protected account that $N'$ corresponds to a subset of *N*.

The *Path Utility Measure*, Figure 3a, computes the average percentage of paths in $G$ that are retained in $G'$. Note that if a node $n$ in $G$ has no corresponding node $n'$ in $G'$, then it contributes 0 to the average.

The nodes in $G$, in Figure 1a, which are not visible via "High-2" are protected by simply hiding the sensitive nodes and their incident edges, yielding the protected account $G_1'$, in Figure 1b. In $G_1'$, the node $b'$ (shown in the figure simply as *b*, since $b' = b$) is only connected to one other node in $G_1'$, while $b$ was connected to 10 nodes in $G$. So, $\%P(b') = 1/10$, while $\%P(h') = 3/10$. Each of the nodes hidden in $G_1'$ contributes 0 to the Path Utility Measure. The Path Utility Measure for $G_1'$ is .13, as shown in Figure 3b. The Path Utility Measures for the graphs in Figure 2 are: (a) .38; (b) .27; (c) .13, the same as for Figure 1c, since the surrogate $f'$ introduces no new paths; (d) .27, the same as for Figure 2a.

But, graph topology is not enough. At the extreme, if you had <null> nodes everywhere, there would be no information in the graph; therefore, we also need some measure of how much of the potential node information is available. This is the Node Utility measure. Node utility is more difficult to assess. Suppose that our social network graph $G$ contained a node with the features <phone, "123-456-7890"> and <name, "Joe">. A surrogate node with the single feature <name, "Joe"> has a lower node utility value than the original node: each node in $G$ is considered maximally useful. Nodes in a protected account are compared to the corresponding original nodes to determine their utility.

The value function $infoScore(n')$, with a range from 0 to 1, reflects the closeness of a node in $G'$ to the corresponding node in $G$, and can depend on completeness, semantic analysis, etc. So, when $n' = n$, $infoScore(n') = 1$. While in general the burden would be too great for administrators to manually assign *infoScores* to all surrogate nodes, we can use defaults or manually assign scores for critically important nodes. The *Node Utility Measure*, Figure 3c computes the average utility of nodes in $G'$ as compared to their corresponding nodes in $G$.

Because the nodes in $G_1'$ were created from $G'$ using an all-or-nothing approach, each node $n'$ in $G_1'$ has an *infoScore* of 1, and the Node Utility Measure of $G_1'$ is 6/11. (Note that the all-or-nothing approach will always yield a Node Utility Measure of $|N'|/|N|$.) In this paper, we also assume that surrogates visible via more restrictive privilege-predicates are more informative. Therefore, the *infoScores* of surrogates defined for the same node are partially ordered: if $n'$ and $n''$ are both surrogates for $n$, and $lowest(n')$ dominates $lowest(n'')$ then $infoScore(n') \geq infoScore(n'')$.

Node and edge surrogate strategies are intertwined. While it may seem that there is a negligible information gain in including a null surrogate, for example, that node may still play an important part in improving the connectivity of the protected account. Conversely, there may be no reason to provide a surrogate for a privilege predicate if none of its incident edges are preserved for that predicate. The Node Utility and Path Utility Measures together help evaluate the utility gains and losses for competing surrogate strategies.

## 4.2 Opacity

A protected account $G'$ is created to satisfy the release policies, so no unauthorized node or edge information is directly revealed to an attacker. However, $G'$ may allow an attacker to infer information in $G$ from $G'$. In other words, *the attacker's goal is to determine the topology of the underlying graph, G, given* $G'$.

A naïve attacker would have no knowledge of the general properties of a graph. Thus, if a naïve attacker is given a disconnected graph, as shown in Figure 1c, the lack of knowledge of the properties of a graph may not awaken any suspicion that the graph has been redacted in any way. On the other hand, an advanced adversary could have fundamental knowledge of the properties of a graph. While an attacker may be able to infer hidden graph information using background domain knowledge of the node's contents, such knowledge is extremely difficult to model and is outside the scope of this paper. Rather we focus on the topology of the graph, and on the attacker's ability to recreate the edges.

Because we wish to protect the protected account from the most advanced attacker, we assume that the attacker has the additional background information:

- Whether or not the complete graph information is represented in $G$,
- The transformations available to the administrator for transforming $G$ into $G'$.

$$opacity(e) = \begin{cases} 0 & \text{if } (n_1' \rightarrow n_2') \in E' \\ 1 & \text{if either } n_1 \text{ or } n_2 \text{ has no corresponding node in } N' \\ 1 - \vartheta & \text{otherwise} \end{cases}$$

$$\vartheta = \frac{\left[\mathrm{FP}(n_1') * \frac{\mathrm{p_C}(n_1',(n_1' \rightarrow n_2'))}{\sum_{i=1}^{j} \mathrm{p_C}(n_1',(n_1' \rightarrow n_i'))}\right] + \left[\mathrm{FP}(n_2') * \frac{\mathrm{p_C}(n_2',(n_1' \rightarrow n_2'))}{\sum_{i=1}^{k} \mathrm{p_C}(n_2',(n_i' \rightarrow n_2'))}\right]}{2}$$

**Figure 4: Opacity Measure**

Using this information, the attacker has a powerful method for determining if the graph presented in the account has been altered. We ignore graph-motif recognition, the possibility that an attacker will be able to recognize sub-graph shapes as corresponding to a particular task. This information is related to an attacker's background knowledge of the graph contents and is outside the scope of this work.

*Opacity* measures the difficulty an attacker is likely to face in inferring the existence of an edge in $G$ that is not present in $G'$. Greater opacity means that an attacker is less likely to infer the existence of the hidden edge. Figure 4 shows the formula for the opacity of an edge $e = (n_1 \rightarrow n_2)$ in $G$. The main calculation, $\vartheta$, relies on two factors: FP, the probability that an attacker will focus on the specific nodes $n_1'$ or $n_2'$ in a protected account, and $p_c(n_1', (n_1' \rightarrow n_2'))$, the likelihood that an attacker focusing on $n_1'$ will infer the edge $(n_1' \rightarrow n_2')$.

For instance, in Figure 1c, because the graph is disconnected and the attacker is advanced, the attacker may focus on the nodes $b'$ and $c'$ as likely nodes with missing edges. There may, of course, be many false positives, but the ability of an attacker to rule these out relies on domain knowledge; for example, in provenance graphs, an attacker knows there aren't any cycles. Background knowledge and expectations determine values for FP. As an example, suppose there is a strong expectation for a strongly connected graph (as would be the case for many provenance queries). As shown in Figure 5 an attacker would be more likely to focus on "loner" nodes, say those connected to at most one other node. Similarly, given that focus, an attacker would be more likely to infer an edge to a node with few edges. Given these sample constants, the table in Table 1 compares path utility and opacity measures for the graphs in Figure 2. Opacity of edge $(f \rightarrow g)$ is 0 for the graph in 2a because the edge is present; it is 1 for 2b because the graph does not include a node corresponding to $f$. Note that when "loner" nodes are likely to draw an attacker's attention, adding surrogate edges to a protected account increase both path utility and opacity (as in graphs 2c and 2d).

The opacity metric is useful in two ways. First, opacity allows an administrator to look at specific nodes and incident edges that are of high security concern and to evaluate the risk of inference. Second, the average opacity over the entire graph can be used to evaluate tradeoffs.

# 5. TRANSFORMATIONS WITH SURROGATES

The utility and opacity measures give us a way to compare protected accounts. In general, the goal is to produce a protected account with as much utility as possible, while making sure that the opacity is still tolerable for the application.

$$\mathrm{FP} = \begin{cases} 0.8, & 0-1 \text{ connected nodes} \\ 0.2, & \text{otherwise} \end{cases}$$

$$p_c = \begin{cases} 0.8, & \text{degree} \leq 1 \\ 0.2, & \text{degree} > 1 \end{cases}$$

**Figure 5: Sample constants for opacity formula for an advanced adversary and an original graph with the properties of i. no disconnected subgraphs and ii. average degree > 1.**

In order to maximize the utility measures, a protected account must: show the original nodes from $G$ whenever possible; include surrogate nodes with the most dominant privilege-predicate available within the high-water set (this is a proxy requirement for choosing the surrogate with the maximal *infoScore*); and reflect as many HW-permitted paths as possible.

*Definition 9. Let $G'$, with high-water set $HW_{G'}$, be a protected account of $G$. $G'$ is maximally informative w.r.t. $G$ iff it satisfies three properties:*

*1. maximal node visibility: For all nodes $n$ in $G$, if $\exists p \in HW_{G'}$ such that $n$ is visible via $p$, then $G'$ includes the corresponding node $n' = n$.*

*2. dominant surrogacy: Let $n$ be a node in $G$ that is not visible via any $p \in HW_{G'}$. If $G'$ contains a corresponding node $n'$, then there is no surrogate $n''$ defined for $n$ such that $n''$ is visible via a privilege-predicate in $HW_{G'}$ and lowest($n''$) dominates lowest($n'$). (Note that if there are incomparable surrogates dominated by $HW_{G'}$, any of them may be chosen for inclusion in $G'$.)*

*3. maximal connectivity: Whenever there is an HW-permitted path between $n_1$ and $n_2$ in $G$, and corresponding nodes $n_1'$ and $n_2'$ appear in $G'$, then there is a path between $n_1'$ and $n_2'$ in $G'$.*

The strength of dominant surrogacy and maximal connectivity depend heavily on the surrogates defined by the provider. When surrogate nodes are provided to take the place of original nodes that are not visible via the high-water set, they allow the inclusion of more surrogate edges, in addition to increasing the node utility.

*Lemma 1*: Let $G'$ be a protected account of $G$, and let $M$ be the subset of nodes in $G$ with corresponding nodes in $G'$. If no node $n \in M$ has surrogates visible via incomparable privilege-predicates in $HW_{G'}$, and $G'$ satisfies the maximum node visibility and dominant surrogacy properties, then no protected account of $G$ with the same set $M$ and the same high-water mark will have a higher node utility measure.

**Table 1: Path Utility and Opacity Measures for graphs in Figure 2.**

| | (a) | (b) | (c) | (d) |
|---|---|---|---|---|
| PathUtility (avg) | .38 | .27 | .13 | .27 |
| Opacity (f→g)only | 0 | 1 | .882 | .948 |

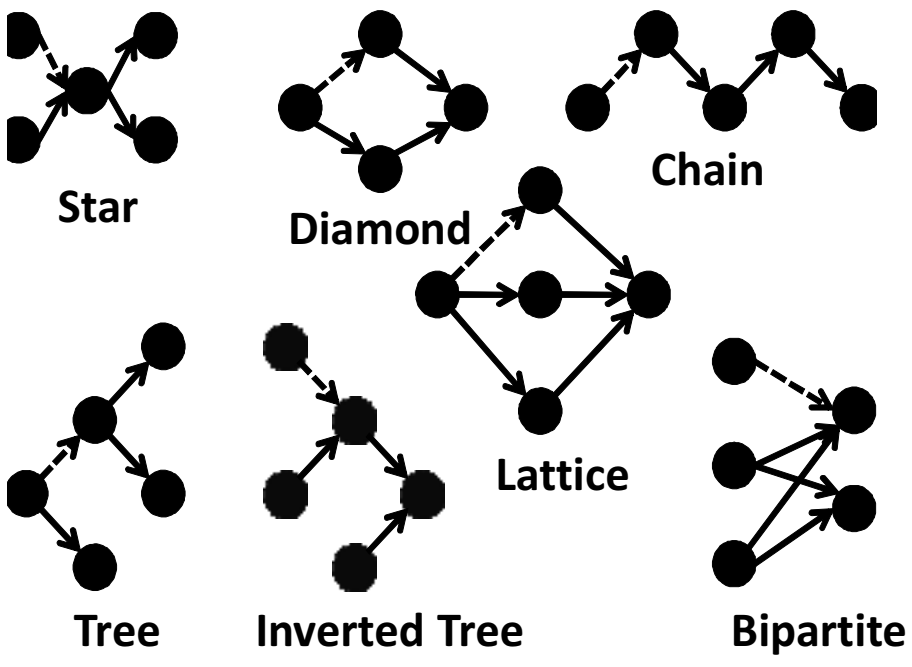

**Figure 6: The Classic Graph Structures used as Motifs. The dashed edge indicates the one chosen to protect.**

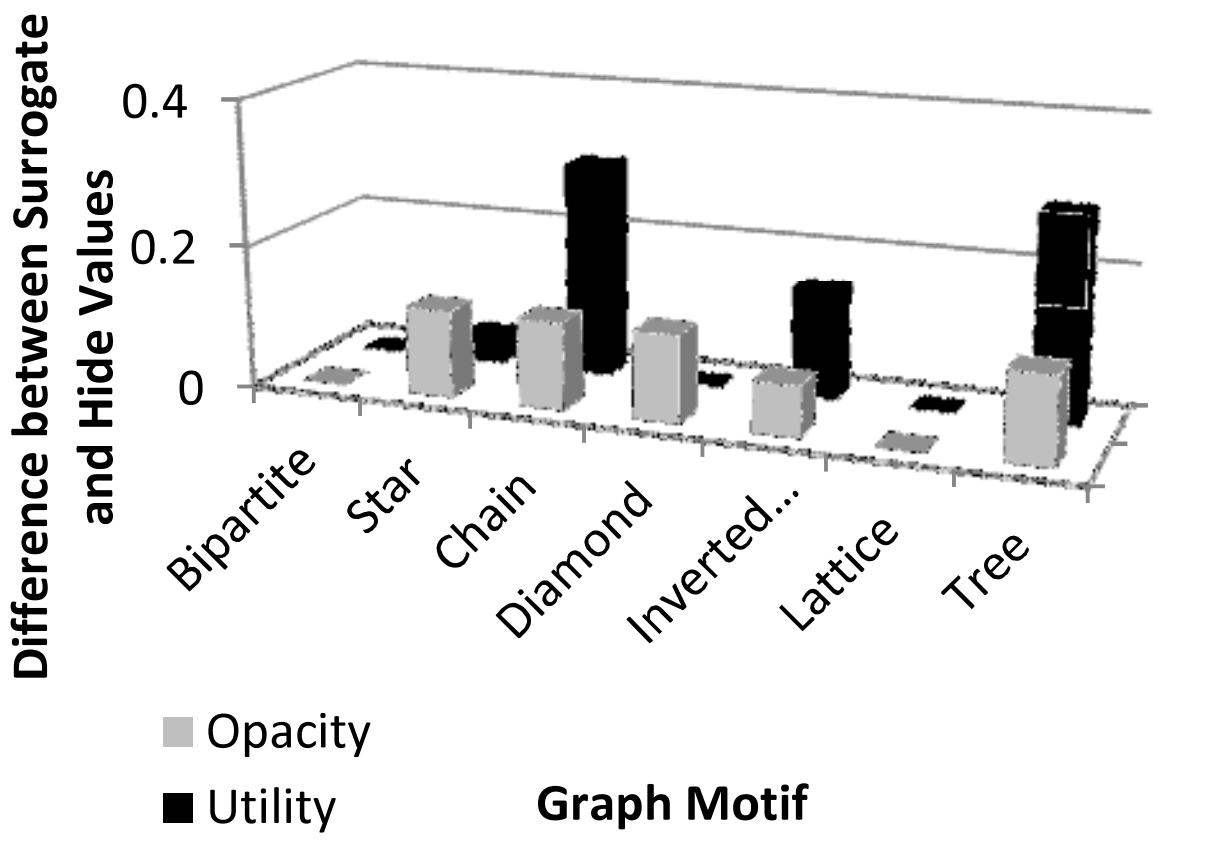

**Figure 7: The difference between surrogating and hiding edges for Utility and Opacity for all motif graphs, e.g. $\text{Opacity}_{\text{Surrogate}}$ - $\text{Opacity}_{\text{Hide}}$.**

When there is a node in *M* with two or more incomparable surrogates, maximum node visibility and dominant surrogacy are sufficient as long as the surrogate with the highest *infoScore* is used in *G'*.

*Lemma 2:* Let *G'* be a protected account of *G*, and let *M* be the subset of nodes in *G* with corresponding nodes in *G'*. If *G'* is maximally connected, then no protected account with the same set *M* and the same high-water mark will have a higher path utility measure.

*THEOREM 1. Suppose that the Surrogate Generation Algorithm generates the protected account G' for a graph G, high-water mark $HW_{G'}$, and set of nodes M with corresponding nodes in G'. Then G' has the highest node and path utility measures possible for $HW_{G'}$ and M .*

The Surrogate Generation Algorithm (see Appendix) is designed to produce maximally informative graphs for the given $HW_{G'}$ and M. Therefore, using the lemmas, the resultant graphs have the highest possible utility measures. The resultant graph is, of course, highly dependent on the high-water marks of the original and protected accounts, the set of node surrogates available, and the policies used to mark node-edge incidences. Different node surrogate and edge marking strategies can be compared via the utility and opacity measures for resulting protected accounts.

# 6. EVALUATION

The question to be answered: is it possible to adequately protect graph information using surrogates while maximizing the usefulness of that provenance information? Our experiments show surrogates give better opacity *and* better utility than removing graph objects.

We perform experiments over two very different sets of graphs: motifs and synthetic. We begin the evaluation of surrogate techniques by studying the effect of surrogates within specialized and well-known graph motifs. Each graph in this set contains only four to five nodes. The nodes are arranged into classic patterns such as tree, diamond, etc. Analysis over these graphs gives insight into how the shape of the graph affects opacity and utility when surrogating or hiding graph information. We then analyze how surrogates affect larger, more complex, synthetically generated graphs. Analysis over these graphs gives insight into how different methods of protecting the graph affect disclosure and utility in more realistic graphs.

While node surrogating is important, we intentionally leave an analysis of node surrogating out of this evaluation. Because node surrogating relies on explicit domain knowledge, it is dependent on specific situations. Edge surrogating can be measured without any assumptions of the attacker's knowledge. The graph transformations were implemented in Java as part of the PLUS system [2]. All experiments were performed on a Dell Windows XP workstation with Intel Dual Core T7400 each at 2.16GHz with 2GB RAM.

## 6.1 Datasets and Experimental Setup

### *6.1.1 Motifs*

We created graphs containing 4-5 nodes and representing classic motifs; these motifs included: star, chain, lattice, diamond, tree, inverted tree and bipartite. Figure 6 contains a representation of all of the motifs utilized. Each motif was used as an initial directed graph, with the first edge protected in each graph, shown as a dashed line.

### *6.1.2 Synthetic Graphs*

While analysis of graph motifs gives insight into the effects of hiding or surrogating edges, typical graphs contain many different motifs. Because these motifs may overlap and interplay differently depending on the links between them, it is important to look at protection in larger graphs. We generated 50 synthetic graphs of 200 nodes each. Across the 50 synthetic graphs, the connectedness increases such that on average every node has 30 to 100 connected pairs; and the protection is varied from 10%-90% of all edges. In other words, for the set of 10 graphs that have 10% of the edges protected, the connected pairs vary from 30 to 100. Additionally, the graphs all contain no disconnected subgraphs and have directed edges

## 6.2 Motif Graph Analysis

Figure 7 shows the change in opacity and utility for all of the motif graphs. Compared to just omitting the information, surrogating raises opacity and utility for all motifs except Bipartite and Lattice. An analysis of these two graphs shows why there is no difference between hiding and surrogating.

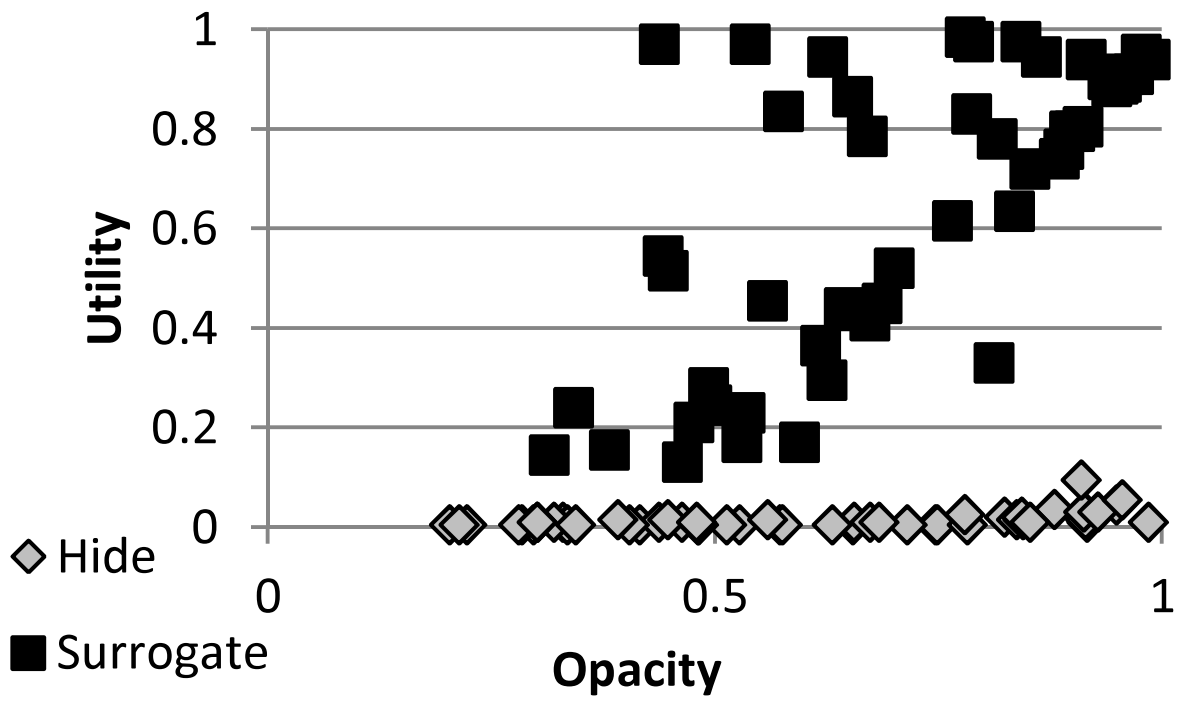

**Figure 8: Maximum Utility given an Opacity rating, as seen in the synthetically generated graphs.**

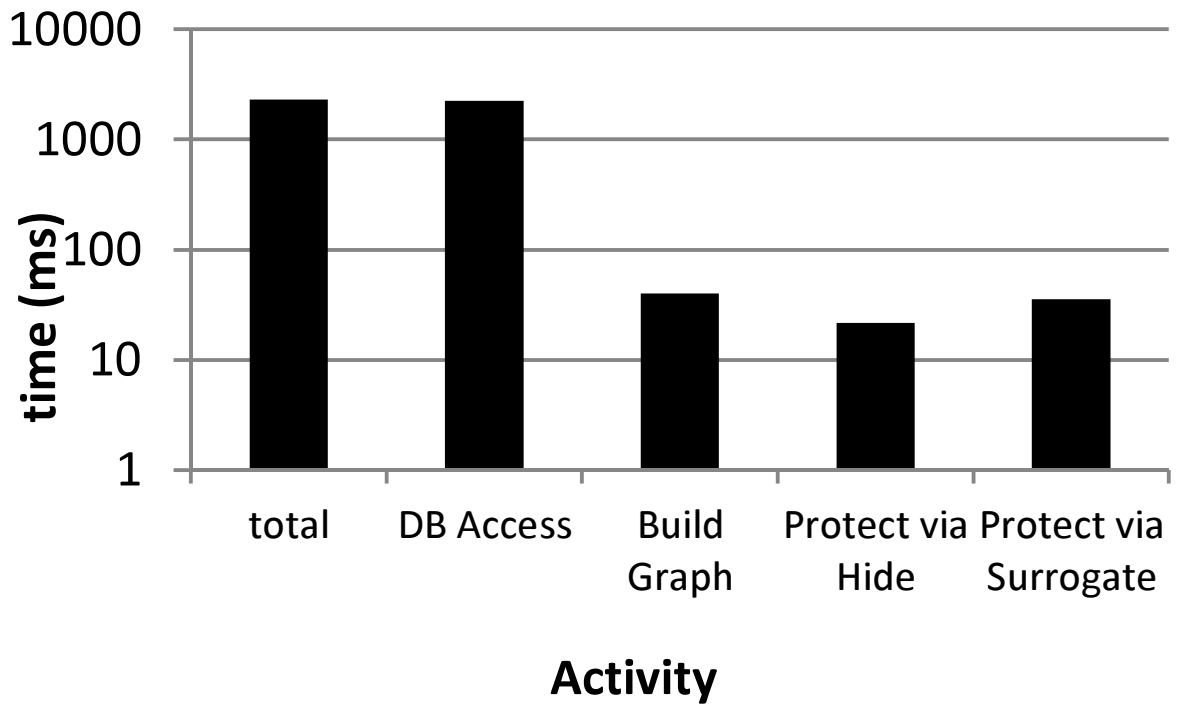

**Figure 10: The time it takes to produce a graph and transform it into a protected account.**

Because the bipartite graph is only two levels deep, the edge that is to be protected cannot be surrogated: there are no nodes in deeper levels that can act as the destination of a surrogate edge, so surrogating in this case produces the same result as hiding. The Lattice graph shows no difference in Utility and Opacity for a different reason. In this case, it is possible to draw a surrogate edge. However, the presence or absence of this edge does not change the utility since the node remains connected to the same set of nodes through other paths. Additionally, the opacity does not change the absence of the edge, in protection by hiding, does not change the Focus Probability (FP), since the degree remains greater than one.

Our experiments show that surrogates can actually increase opacity, making it harder for an adversary. This is because edges summarizing paths are created, lowering the suspicion of a node without edges .

## 6.3 Synthetic Graph Analysis

Figure 9 shows the difference in opacity between surrogating vs. hiding edges in larger synthetic graphs. We use the graph notion of connected pairs to measure how interconnected or sparse a graph is. The amount of the graph protected has the greatest influence on how much better surrogating is than hiding. Figure 9a shows the difference in Opacity. The more surrogates used for protection, the greater the opacity. Meanwhile, Figure 9b shows that the utility goes down based on the amount of graph that is protected. The key takeaway, however, is that all of the values in Figure 9 are positive; surrogating is always better than hiding for protection.

Figure 8 shows the tradeoff between opacity and utility for the hiding vs. surrogating decisions. Even if an administrator wishes to protect the graph and ensure a particular opacity, it is better to use surrogates to maintain a desired opacity while sharing more useful graphs.

## 6.4 Performance

Figure 10 shows the time it takes to create the base account and the protected account. The transformation from $G$ to $G'$ takes ~10ms. Hiding takes less time since the overall size of the graph is ultimately smaller. Thus, the cost for protecting a graph via transformation into a protected account is easily subsumed in the cost of creation of the graph itself.

# 7. RELATED WORK

As noted in the Introduction, there is considerable research on whole-graph manipulations of social network graphs to protect privacy while supporting data analysis [1, 5, 9, 10, 15, 16]. There is also substantial research on access controls for provenance [3, 4, 8, 12-14]. None of this work addresses the problem of uninformative query results for path-traversal queries.

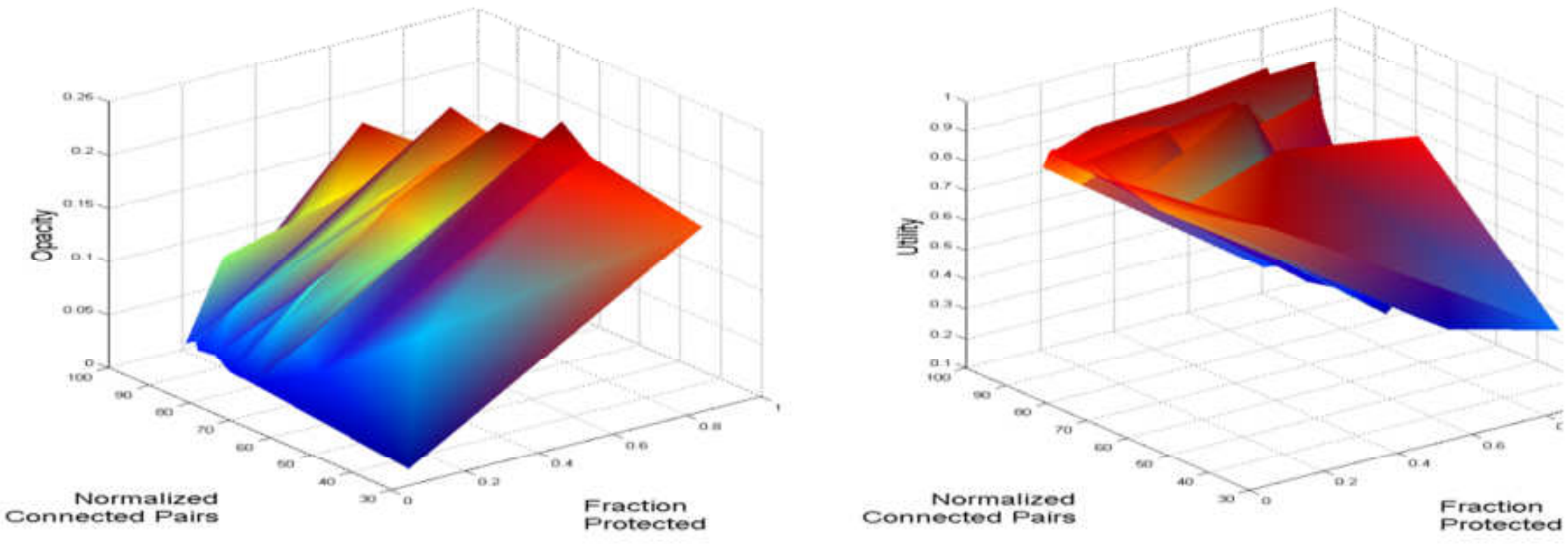

**Figure 9: The difference between surrogating and hiding edges for Utility and Opacity for all synthetic graphs, e.g. $\text{Opacity}_{\text{Surrogate}}$ - $\text{Opacity}_{\text{Hide}}$. The graphs show how Utility and Opacity change with hiding and surrogating based on how connected a graph is, and how much of it is protected.**

[7] uses views to hide details of how modules are connected in a workflow. Also, subsetting approaches such as Oracle Label Security automatically define a view that does not access the sensitive information, while predefined views allow for arbitrary surrogates. These view-based approaches suffer the major disadvantages of: manual view definition for all views (requiring detailed knowledge of a subgraph and all its potential user classes), and view recomputation when object sensitivity changes. In contrast, surrogates and protected accounts allow owners of particular nodes to locally specify their security concerns and have the appropriate views constructed automatically. On the other hand, techniques such as [6] can be used in parallel with surrogates, by using the Secure-View to suggest appropriate modules and attributes to hide.

# 8. CONCLUSIONS

In this work, we describe the need to protect graph information beyond complete node or edge anonymization. We show that there are graphs in which only certain nodes and edges must also be protected, and that a binary “show/hide” approach is not generally useful. To this end, we introduce the notion of surrogates that can protect the underlying graph information by providing less detailed node and edge information. Moreover, we describe metrics to measure how much an attacker may be able to learn about the underlying graph based on the protected account generated. We also introduce a measure for the usefulness of the protected account with respect to the original graph. Through experiments on small classic graphs, and large synthetic graphs, we show that our surrogating approach generally provides better utility and opacity than the “show/hide” approach.

# 9. ACKNOWLEDGEMENTS

We thank the Principles of Provenance Symposium 2009 participants: Peter Buneman, James Cheney, Ashish Gehani, Michael Hicks, Kristen LeFevre, Andrew Martin, Gerome Miklau and Margo Seltzer for assistance in clarifying the problem statement. We are grateful to the anonymous PVLDB reviewers for their incisive questions and suggestions. Finally, we thank the MITRE Innovation Program for funding support.

# 10. REFERENCES

[1] L. Backstrom, C. Dwork, and J. Kleinberg, "Wherefore Art Thou R3579X? Anonymized Social Networks, Hidden Patterns, and Structural Steganography," *Proc. 16th Intl. World Wide Web Conference*, pp. 181-190, 2007.

[2] B. T. Blaustein, L. Seligman, M. Morse, M. D. Allen, and A. Rosenthal, "PLUS: Synthesizing privacy, lineage, uncertainty and security," *ICDE Workshops*, pp. 242-245, 2008.

[3] U. Braun, A. Shinnar, and M. Seltzer, "Securing Provenance," *USENIX HotSec*, pp. 1-5, 2008.

[4] A. Chebotko, S. Chang, S. Lu, F. Fotouhi, and P. Yang, "Scientific Workflow Provenance Querying with Security Views," *WAIM*, pp. 349-356, 2008.

[5] G. Cormode, D. Srivastava, S. Bhagat, and B. Krishnamurthy, "Class-based graph anonymization for social network data," *PVLDB*, vol. 2, pp. 766-777, 2009.

[6] S. Davidson, S. Khanna, T. Milo, D. Panigrahi, and S. Roy, "Provenance Views for Module Privacy," in *PODS*, 2011.

[7] S. B. Davidson, S. Khanna, V. Tannen, S. Roy, Y. Chen, T. Milo, and J. Stoyanovich, "Enabling Privacy in Provenance-Aware Workflow Systems," *CIDR*, pp. 215-218, 2011.

[8] R. Hasan, R. Sion, and M. Winslett, "Introducing Secure Provenance: Problems and Challenges," *Proceedings of the 3rd International Workshop on Storage Security and Survivability*, pp. 13-18, 2007.

[9] M. Hay, G. Miklau, D. Jensen, D. Towsley, and P. Weis, "Resisting Structural Identification in Anonymized Social Networks," *VLDB*, pp. 102-114, 2008.

[10] K. Liu and E. Terzi, "Towards Identity Anonymization on Graphs," *SIGMOD*, pp. 93-106, 2008.

[11] L. Moreau, B. Clifford, J. Freire, J. Futrelle, Y. Gil, P. Groth, N. Kwasnikowska, S. Miles, P. Missier, J. Myers, B. Plale, Y. Simmhan, E. Stephan, and J. Van den Bussche, "The Open Provenance Model core specification (v1.1)," *Future Generation Computer Systems*, 2010.

[12] A. Rosenthal, L. Seligman, A. Chapman, and B. Blaustein, "Scalable Access Controls for Lineage," *First Workshop on Theory and Practice of Provenance Systems (TaPP)*, 2009.

[13] V. Tan, P. Groth, S. Miles, S. Jiang, S. Munroe, S. Tsasakou, and L. Moreau, "Security Issues in a SOA-Based Provenance System," pp. 203-211, 2006.

[14] J. Zhang, A. Chapman, and K. LeFevre, "Fine-Grained Tamper-Evident Data Pedigree," *Secure Data Management*, pp. 17-32, 2009.

[15] E. Zheleva and L. Getoor, "Preserving the Privacy of Sensitive Relationships in Graph Data," *PinKDD*, pp. 153-171, 2007.

[16] B. Zhou and J. Pei, "Preserving Privacy in Social Networks Against Neighborhood Attacks," *ICDE*, pp. 506-515, 2008.

# APPENDIX

## A: Provenance Example

To prepare for a disease outbreak, an emergency treatment plan is produced via the set of processes described in Figure 11a, with arrows indicating flow over time. The shading of each node indicates the lowest of the privilege classes defined in Figure 11b required to access the node. Arrows in Figure 11b indicate a "dominates" relationship, e.g., Cleared Emergency Responder (CER) dominates Emergency Responder (ER), meaning that all users with CER privileges can access all nodes marked with ER. A user with ER privileges can see the Emergency Treatment Plan, but not information on the Local Action Planning process or the Emergency Supplies Stockpiles data which fed it. In prior provenance systems, that same ER user could not learn anything about the provenance of the Emergency Treatment Plan, even though she has sufficient privileges to access many of the contributing nodes if only she knew they existed.

## B: Producing Protected Accounts

Algorithms 1-3 in Figure 12 contain pseudocode for generating maximally informative protected accounts. For simplicity, the algorithm shown here considers protected accounts with singleton high-water sets; when there are multiple privilege-predicates, the same process is used for each predicate until an appropriate surrogate is found. When surrogates are available for incomparable privilege-predicates, we assume there is a domain-dependent function to choose which surrogate is included in the protected account.)

We start with an original graph, $G = (N, E)$, where bi-directional edges are modeled as two directional edges, and the algorithm produces a protected account $G' = (N', E')$ such that HW$_{G'}$ = {$p$} and $p$ is dominated by some member of HW$_G$. This protected account can be used to evaluate queries by consumers satisfying $p$. Whenever a node in $G$ is visible via p, that node is included in $N'$ (upholding the maximal node visibility property), otherwise the visible surrogate with the highest info-score is used. The edge markings are used to determine $E'$. In Algorithm 2, every edge $(n_1, n_2) \in E$ which is marked Visible at privilege $p$ by both $n_1$ and $n_2$ is preserved in $E'$. Edges marked Hidden by either the source or destination node are excluded from $E'$. Other edges, i.e., those with no Hide markings, and with at lowest one Surrogate marking, are used to compute surrogate edges in accordance with the maximal connectivity property. In particular, $E'$ includes an edge from $n'_1$ to $n'_2$ iff:

1. $(n_1, n_2)$ has no Hide marking for $p$ (i.e., if $E$ included $(n_1, n_2)$, then we don't want to show any computed edges between them either); the rest of the conditions allow for replacing paths with surrogate edges

2. there is an HW$_G$-permitted path from $n_1$ to $n_2$ in $G$, and

3. there is no shorter HW$_G$-permitted path from $n_1$ to $n_2$ in $G$ (redundant surrogate edges would not violate any correctness criteria, but they make the graph less clear; also note that paths consisting only of edges marked Visible will not yield surrogate edges because each preserved edge is shorter than the path, and these are already included in $E'$)

This algorithm runs in $O(n^2d)$ where $n$ is the number of nodes in the graph, $d$ is the degree of the nodes. Every node will be touched once to determine a surrogate. However, when determining the surrogate edges, the worst case is when all edges are marked Surrogate. In this case, every node will look forwards and backwards through its edges attempting to find a visible node to make an edge with, thus possibly touching n nodes in a fully connected graph.

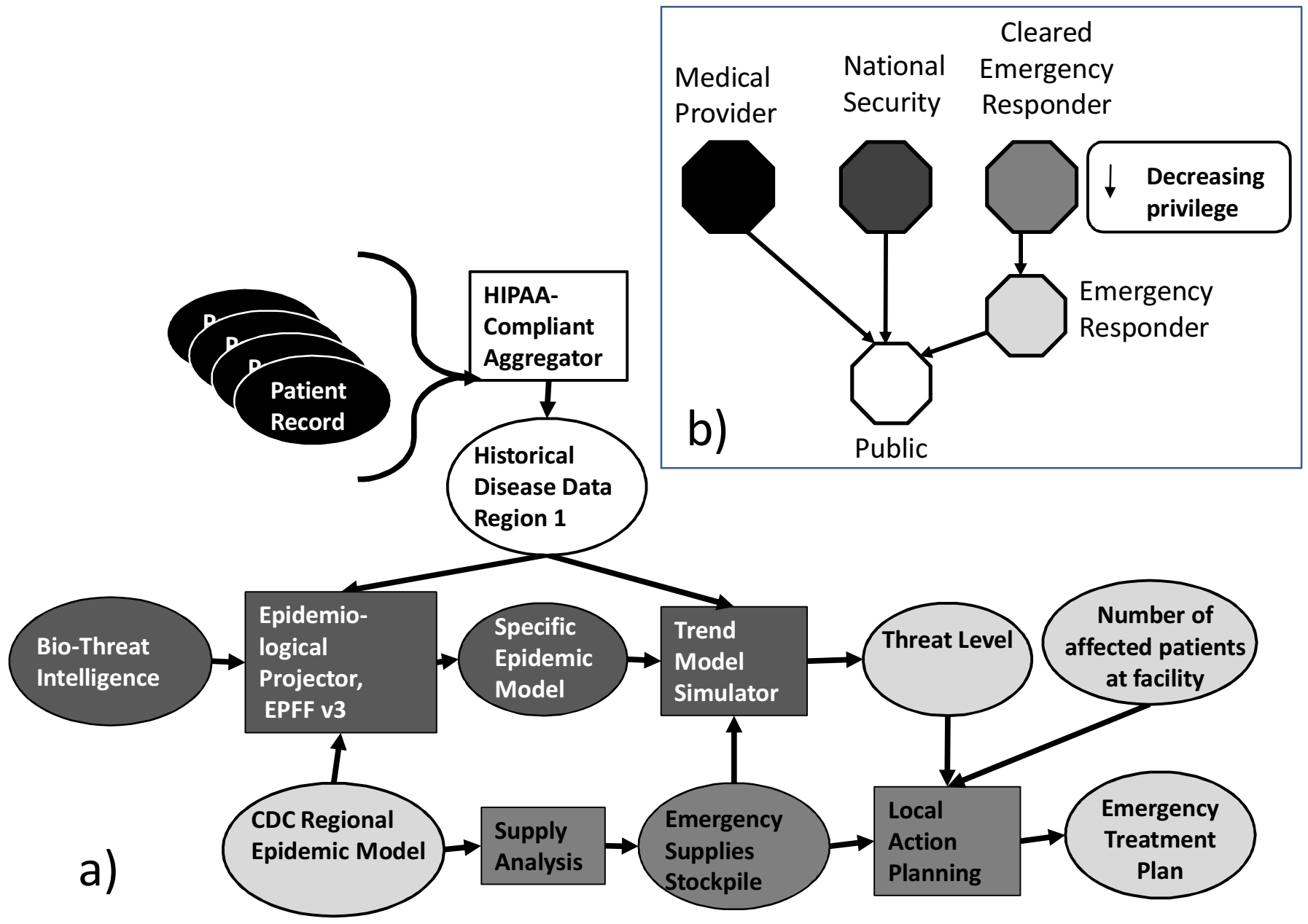

**Figure 11 Example of a provenance graph to be protected, similar to the generic graph found in Figure 1.**

**Algorithm 1: Protected Account Generation**

```
Input: Graph G =(N, E).
Input: Privilege-predicate p.
Output: Protected Account G' =(N', E').
for Nodes n ∈ N do
    if p dominates lowest(n) then
        N'.add(n)
    else
        N'.add(most-dominant-surrogate(p,n))
    end if
end for
Set potentialEdges = markEdges(G, p)
for Edges <n1, n2> ∈ potentialEdges do
    if marking = Visible then
        E'.add(<n1, n2>)
    else
        Node backwards; Node forwards;
        if edge-mark(p, n1) = Visible & edge--mark(p,n2) = Surrogate then
            backwards=n2; forwards=n1;
        else if edge-mark(p, n1) = Surrogate & edge-mark(p,n2) = Visible then
            backwards=n2; forwards=n2;
        else if edge-mark(p, n1) = Surrogate & edge-mark(p,n2) = Surrogate then
            backwards=n1; forwards=n2;
        end if
        Set backset = buildVisibleSet(backwards, potentialEdges, backwards)
        Set forwardset= buildVisibleSet(forwards, potentialEdges, forwards)
        for Node pairs (f, b), f ∈ forwardset, b ∈ backset do
            N'.add(<f,b>);
        end for
    end if
end for
```

**Algorithm 2: Build Visible Set.**

```
Input: Node n
Input Privilege Predicate p.
Output: Set reachableNodes
for Edges <n, n1> ∈ E do
    if edge-mark(p, n1) = VISIBLE then
        : reachableNodes.add(other)
    else reachableNodes.add(BuildVisibleSet(,n1, p))
    end if
end for
return reachableNodes
```

**Algorithm 3: MarkEdges**

```
Input: Graph G =(N, E).
Input: Privilege Predicate p.
Output: Set potentialEdges
for Edges <n1, n2>  ∈ E do
    if edge-mark(p, n1) = Visible & edge-mark(p,n2) = Visible then
        potentialEdges.add([<n1, n2>, Visible]);
    else if not edge-mark(p, n1) = Hide & not edge-mark(p,n2) = Hide then
        potentialEdges.add([<n1, n2>, Surrogate]);
    end if
end for
return potentialEdges
```

**Figure 12: Surrogate Generation Algorithms**